\documentclass[
superscriptaddress,
amsmath,amssymb,
 aps,
prl,
twocolumn,
floatfix,
longbibliography
]{revtex4-2} 
\usepackage[T1]{fontenc} 
\usepackage{graphicx}
\usepackage{bm}
\usepackage{color}
\usepackage[hyphen]{xurl}
\usepackage{appendix}
\usepackage{caption, subcaption} 
\usepackage{multirow}
\begin{document}
\title{Multifractional Brownian motion with telegraphic, stochastically varying exponent
}

\author{Michał Balcerek}
\affiliation{Faculty of Pure and Applied Mathematics, Hugo Steinhaus Center, Wrocław University of Science and Technology, Wybrzeże Wyspiańskiego 27, 50-370 Wrocław, Poland}

\author{Samudrajit Thapa}
\email{thapa@pks.mpg.de}
 \affiliation{Max Planck Institute for the Physics of Complex Systems, Noethnitzer Straße 38, 01187 Dresden, Germany }
\affiliation{Department of Physics, Indian Institute of Technology Guwahati, Guwahati 781039, Assam, India}
 \author{Krzysztof Burnecki}
 \affiliation{Faculty of Pure and Applied Mathematics, Hugo Steinhaus Center, Wrocław University of Science and Technology, Wybrzeże Wyspiańskiego 27, 50-370 Wrocław, Poland}
 
\author{Holger Kantz}
\affiliation{Max Planck Institute for the Physics of Complex Systems, Noethnitzer Straße 38, 01187 Dresden, Germany }
 
\author{Ralf Metzler}%
  \affiliation{Institute of Physics \& Astronomy, University of Potsdam,   14476 Potsdam, Germany}
  
\author{Agnieszka Wyłomańska}
\affiliation{Faculty of Pure and Applied Mathematics, Hugo Steinhaus Center, Wrocław University of Science and Technology, Wybrzeże Wyspiańskiego 27, 50-370 Wrocław, Poland}%
 
\author{Aleksei Chechkin}%
  \affiliation{Faculty of Pure and Applied Mathematics, Hugo Steinhaus Center, Wrocław University of Science and Technology, Wybrzeże Wyspiańskiego 27, 50-370 Wrocław, Poland}
  \affiliation{Institute of Physics \& Astronomy, University of Potsdam,   14476 Potsdam, Germany}
 \affiliation{German-Ukrainian Core of Excellence, Max Planck Institute of Microstructure Physics, \\ Weinberg 2,  06120 Halle, Germany}
 \affiliation{Akhiezer Institute for Theoretical Physics, National Science Center ‘Kharkiv Institute of Physics and Technology’, Akademichna
st.1, Kharkiv 61108, Ukraine} 
  
\date{\today}             
\begin{abstract}
The diversity
of diffusive systems exhibiting long-range correlations characterized by a stochastically varying Hurst exponent calls for a generic multifractional model. We present a simple, analytically tractable model which fills the gap between mathematical formulations of  multifractional Brownian motion and empirical studies. In our model, called telegraphic multifractional Brownian motion, the Hurst exponent is modelled by a  smoothed telegraph process which results in a stationary beta distribution of exponents as observed in biological experiments. We also provide a methodology to identify our model in experimental data and present concrete examples from biology, climate and finance to demonstrate the efficacy of our approach.  \end{abstract}

\maketitle

When creating the mathematical basis of the theory of locally homogeneous isotropic turbulence~\cite{frisch95}, Kolmogorov introduced a new class of random processes called \textit{``Wiener spirals''}, which are Gaussian self-similar processes with stationary, power-law correlated increments~\cite{kol40a,kol40b}. The theory of Wiener spirals have received further development in mathematical literature \cite{yaglom1953random,yaglom1955,yaglom2012}, however, remained almost unknown to a broader scientific community until 1968, when Mandelbrot and van Ness~\cite{NessMandelbrot} presented an explicit integral representation for such process that they called \textit{``fractional Brownian motion''} (FBM). FBM has become a paradigmatic example of a Gaussian random process whose scaling properties are characterized by a unique index $H$ called \textit{Hurst exponent} ($0<H<1$)~\cite{mishura2008stochastic}. It exhibits persistent behavior (i.e. supporting the existing tendency) when $1/2< H < 1$, and antipersistent behavior (supporting the opposite tendency) when $0<H<1/2$. In anomalous diffusion theory the diffusion exponent characterizing the time dependence of the mean squared displacement (MSD) for FBM equals $2H$, thus reflecting either fast (super-), or slow (sub-) diffusion for $H>1/2$ and $H<1/2$, respectively~\cite{alla2019stochastic,sokolov12,franosch2013,metzler2014}. The case $H=1/2$ corresponds to ordinary Brownian motion, or the Wiener process $B(t)$~\cite{alla2019stochastic}.   Numerous phenomena exhibiting FBM-like behavior were found in diverse fields, from telecommunications, engineering~\cite{Dou18,dekking99} and image processing~\cite{jennane01} to astrophysics~\cite{miville2003}, climate~\cite{metzler2021largedev,janosi2022}, underground water transport~\cite{book:korvin}, from movement ecology~\cite{vilk2022,vilk2022unravelling}, intracellular motion \cite{franosch2013,weiss2012, krapf2019spectral} and paths of nerve fibres~\cite{januvsonis2020} to financial mathematics~\cite{mishura2008stochastic,FBMvolatility}. 

However, both hallmarks of FBM, power-law correlations and self-similarity, imply strong idealizations which often are not realized in practice. As a generalization of FBM, \textit{Multifractional Brownian motion} (MBM) was introduced in the mathematical literature: a random process characterized by a function $H(t)$ which can be either deterministic or random~\cite{levy1995multifractional,benassi1997elliptic,cohen1999self,ayache2000covariance,ayachetaqqu2005,stoev2006rich,ryv15,ayache2010process}.
Indeed, in many experimental observations there is evidence that the Hurst exponent $H$ randomly varies from realization to realization or even along a single sample path. Such doubly stochastic behavior was observed in financial data~\cite{bianchi2013modeling, bianchi2014multifractional, Tapiero_Risk2016}, segmentation of images~\cite{combrexelle2015}, pollution data~\cite{mastalerz2018pollution}, and recently in several single-particle tracking (SPT) experiments~\cite{Pawar_2014,Schwiezer_2015,Stiehl_2016,wagner2017,wang2018,cherstvy2018non,cherstvy2019non,vacuoles2019,sabri2020elucidating,han2020deciphering,benelli2021sub,speckner2021single,janczura2021identifying,Perkins_2021,cherstvy2018non,korabel2021local,waigh2023heterogeneous}.
Therefore, it is likely a generic feature of a certain class of systems. However, to the best of our knowledge, there is an absence of generic analytical models of MBM to compare with the empirical observations.

In this Letter we aim at filling the gap between mathematical formulations of MBM with random Hurst exponent and experiments by introducing \textit{telegraphic multifractional Brownian motion} (TeMBM),
a simple generic analytical model mimicking smooth variations of the Hurst exponent along the sample path.
We provide a methodology to distinguish between three classes of power-law correlated random processes, namely FBM (with fixed Hurst exponent), FBM with Hurst exponent varying between different realizations, and TeMBM.
We present examples from biology, climate and finance to demonstrate the efficacy and applicability of our approach.

\begin{figure*}[t] 
\begin{subfigure}{.45\textwidth}
 \centering
  \includegraphics[width=.9\linewidth]{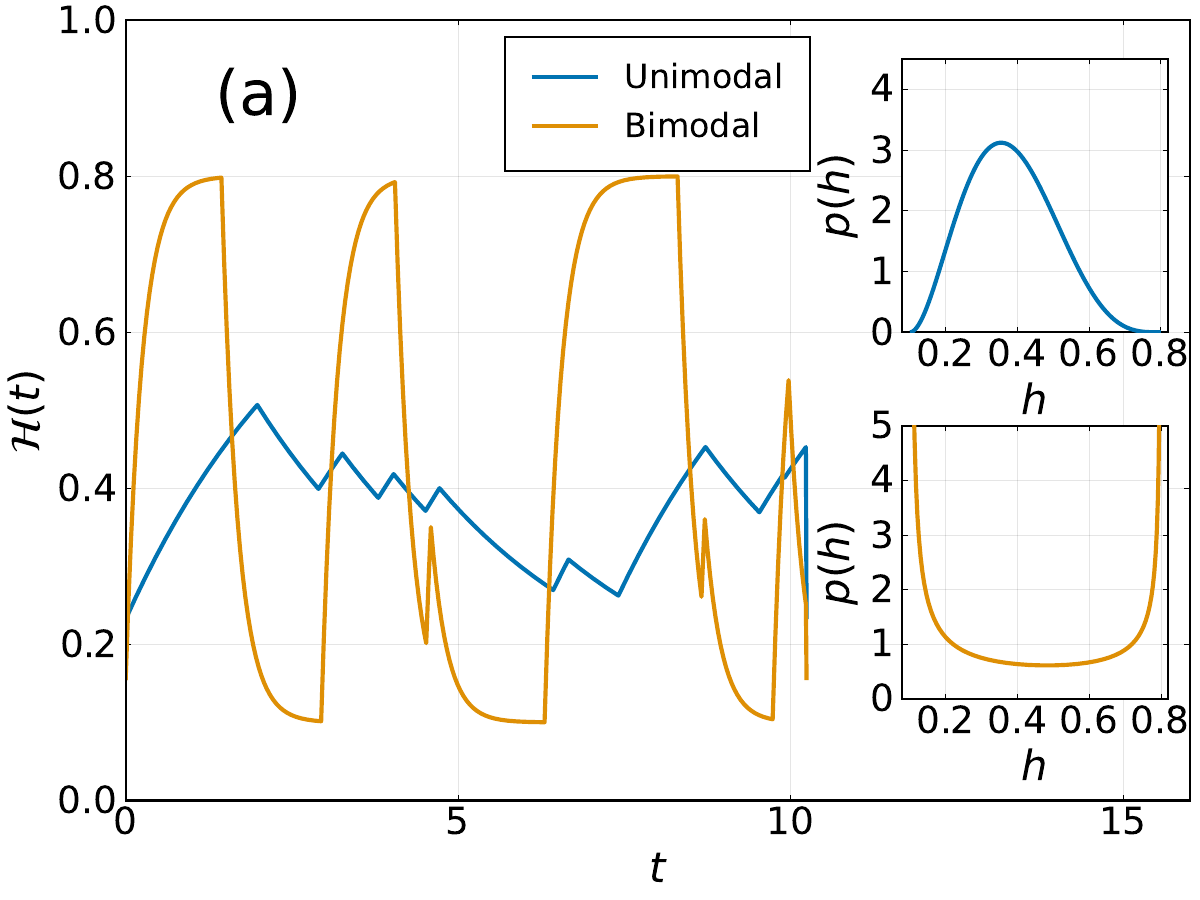}  
  \vskip -.2cm
\end{subfigure}
\begin{subfigure}{.45\textwidth}
  \centering
  \includegraphics[width=.9\linewidth]{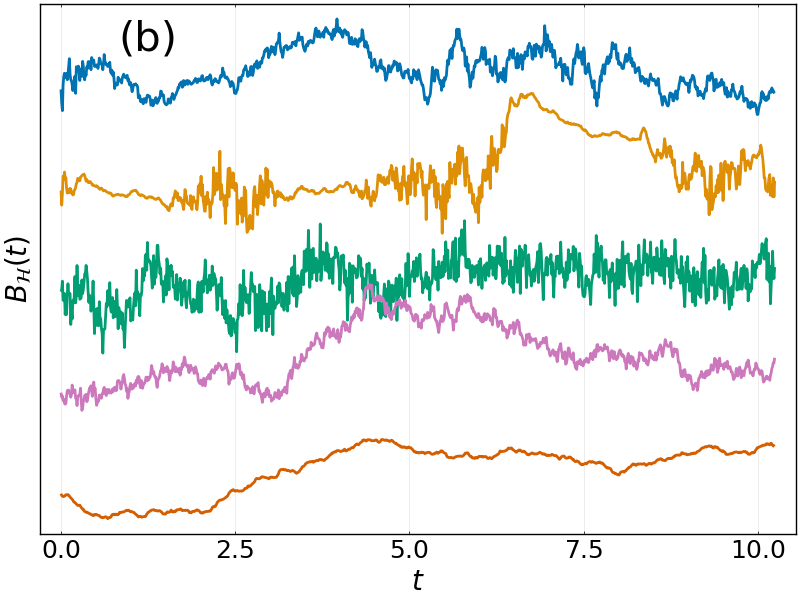}
  \vskip -.2cm
\end{subfigure}
\caption{(a) Single trajectories of $\mathcal{H}(t)$ with the following parameters: $\lambda_{12} = 1, \lambda_{21} =1.5, \tau = 3$ (unimodal), $\tau = \frac{1}{4}$ (bimodal), and the levels $H_1 = 0.1, H_2 = 0.8$. Insets: the corresponding PDFs given by Eq.~(\ref{eq-beta-dist}). (b) Single trajectories corresponding to (from top to bottom) TeMBM (unimodal case), TeMBM (bimodal case), FBM with $H=0.1$, FBM with $H=0.4$, and FBM with $H=0.8$. The individual trajectories are shifted with respect to each other for better visibility.  }
\label{fig:1}
\end{figure*}

We define MBM with random Hurst exponent via  the spectral representation~\cite{benassi1997elliptic,stoev2006rich,cohen1999self,ayache2000covariance}
\begin{align}
    B_{\mathcal{H}}(t) = C(\mathcal{H}(t))\int_{-\infty}^{\infty} \frac{e^{\mathrm{i}\omega t}-1}{|\omega|^{\mathcal{H}(t)+1/2}} dB(\omega), \quad t\geq 0,
\end{align}
where $dB(\omega)$ is "the Fourier transform" of the white noise  $dB(t)$ with $\langle dB(\omega_1) dB(\omega_2) \rangle = \delta(\omega_1+\omega_2) d\omega_1 d\omega_2$ \cite{ayache2010process}, and $\mathcal{H}(t)$ is a stationary process which is independent of  $B(t)$. Its probability density function (PDF) $p(h)$ is defined on the interval $0 \leq h \leq 1$. The prefactor $C(\mathcal{H}(t))=\sqrt{\Gamma(2\mathcal{H}(t)+1) \sin(\pi \mathcal{H}(t))}/\sqrt{2\pi}$ is chosen such that the MSD conditional on $\mathcal{H}(t)$ is $\left\langle ( B_{\mathcal{H}}(t) - B_{\mathcal{H}}(0) )^2 \right\rangle = t^{2\mathcal{H}(t)}$.

The autocovariance function (ACVF) of $B_{\mathcal{H}(t)}$ conditional on $\mathcal{H}(t)$ thus takes the form 
\begin{eqnarray}
\label{eq-fbm-autocorr}
    \langle B_{\mathcal{H}}(t) B_{\mathcal{H}}(s) \rangle &=& D(\mathcal{H}(t), \mathcal{H}(s)) \left(t^{\mathcal{H}(t)+\mathcal{H}(s)} + s^{\mathcal{H}(t)+\mathcal{H}(s)} \right)\nonumber\\
    &-&D(\mathcal{H}(t), \mathcal{H}(s))|t-s|^{\mathcal{H}(t)+\mathcal{H}(s)},
\end{eqnarray}
where the function $D(x,y)$ is defined as (see Section I in Supplemental Material~\cite{supp}~\nocite{sancho1984,shapiro1978,skorokhod1961stochastic,giorno1986some,ward2003diffusion,web:source} for more details)\begin{eqnarray}D(x, y)=\frac{\sqrt{\Gamma(2x+1)\Gamma(2y+1)\sin(\pi x) \sin (\pi y)}}{2 \Gamma(x+y+1) \sin (\frac{\pi(x+y)}{2})}.\end{eqnarray}  Note that if $\mathcal{H}(t)$ is constant, Eq.~(\ref{eq-fbm-autocorr}) reduces to the well-known ACVF which uniquely defines the Kolmogorov-Mandelbrot FBM. The PDF of  $B_\mathcal{H}(t)$ is given by
\begin{align}
  P(x,t)= \frac{1}{\sqrt{2\pi}}\int_{-\infty}^{\infty} \frac{\exp\left\{-x^2/(2t^{2h}) \right\}}{ t^{h}} p(h) dh,
  \label{eq-form-pdf}
\end{align}
where $p(h)$ is defined below.

\begin{figure*}[t] 
\begin{subfigure}{.45\textwidth}
  \centering
  \includegraphics[width=.9\linewidth]{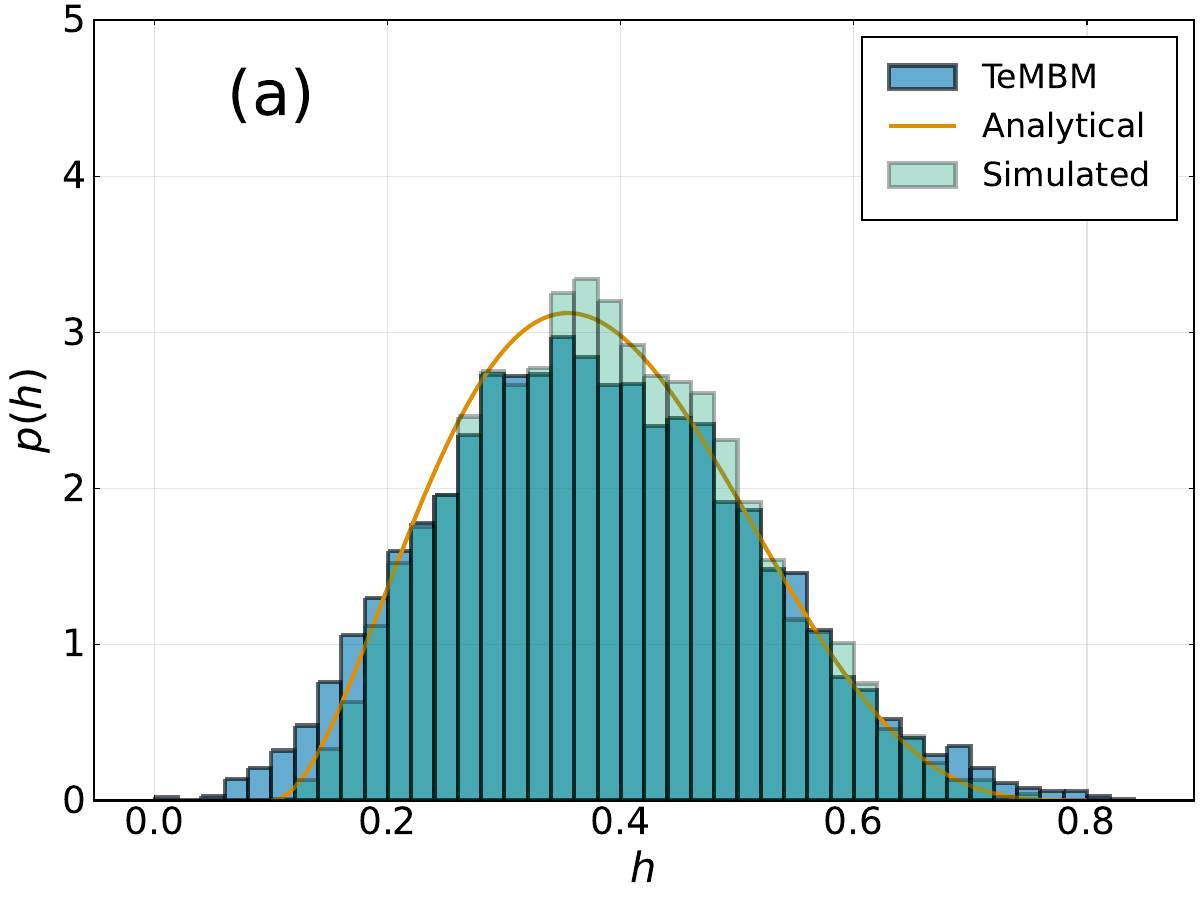}  
  \vskip -.2cm
\end{subfigure}
\begin{subfigure}{.45\textwidth}
  \centering
  \includegraphics[width=.9\linewidth]{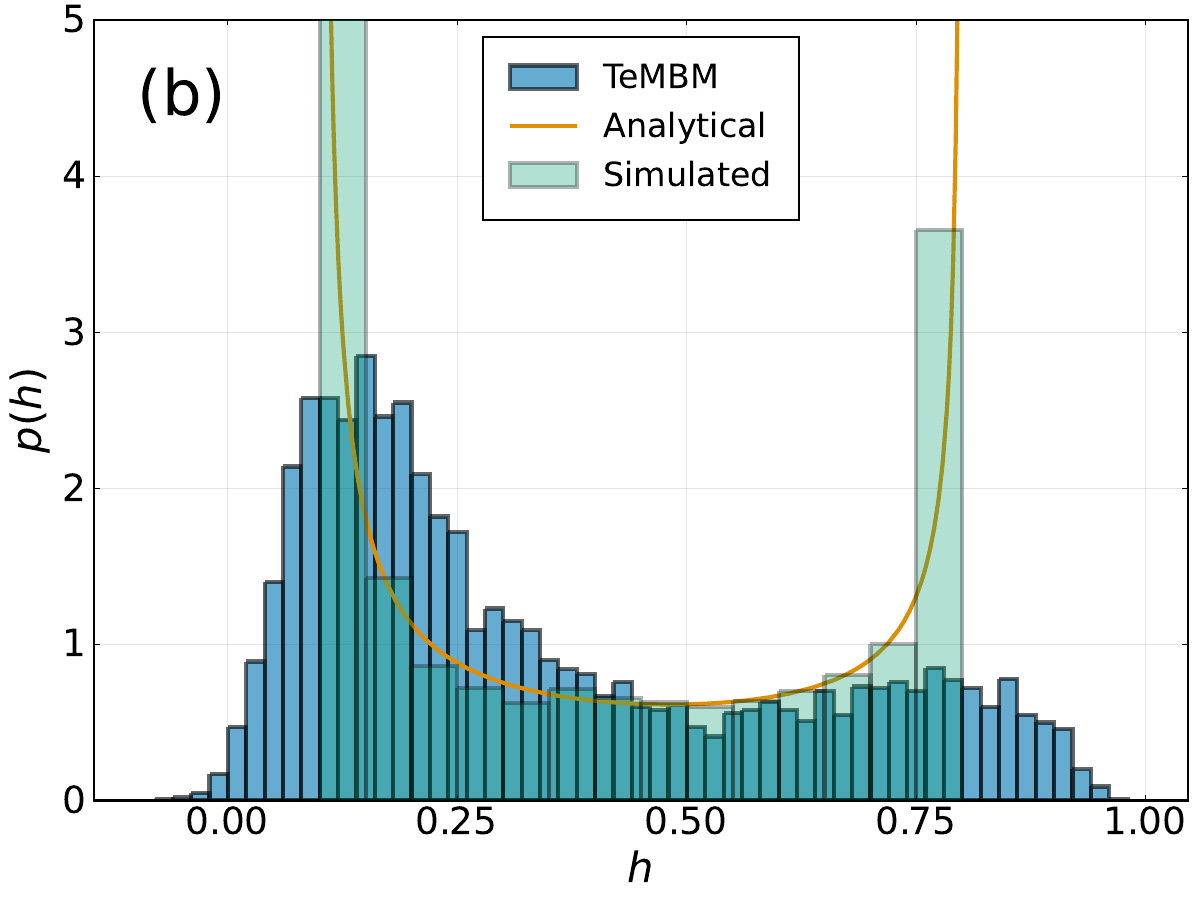}
  \vskip -.2cm
\end{subfigure}
\caption{PDFs of estimated values of the Hurst exponent for (a) unimodal case, (b) bimodal case. The parameters are the same as used in Fig.~\ref{fig:1}. See text for details.}
\label{fig:2}
\end{figure*}

To go further we need to specify the process $\mathcal{H}(t)$. Here we propose to model the temporal variability of the Hurst exponent with a  stationary smoothed telegraph process defined through the stochastic differential equation
\begin{align}
\frac{d\mathcal{H}_{}(t)}{dt}=\frac{-\mathcal{H}_{}(t)+\mathcal{H}_{TP}(t)}{\tau},
\label{eq:stp-def}
\end{align}
where $\tau$ is the relaxation time and $\mathcal{H}_{TP}(t)$ is a telegraph process, i.e., a stationary dichotomic Markov process that jumps between the two values $H_1$ and $H_2$, $0\leq H_1 <H_2 \leq 1$,  with mean rates $\lambda(H_1 \rightarrow H_2)=\lambda_{12}$ and 
$\lambda(H_2 \rightarrow H_1)=\lambda_{21}$. We call the resulting process $B_{\mathcal{H}}(t)$  TeMBM.  We note that the telegraph process and its extensions are useful to model financial market dynamics~\cite{tp_finance1,tp_finance2,tp_finance3,tp_finance4} as well as the dynamics in biological systems, for example, gene expression~\cite{paulsson2005models,tsimring2014noise} or when myosin motors exert contractile forces on the cytoskeleton network~\cite{mizuno2007nonequilibrium,mackintosh2008nonequilibrium,brangwynne2008nonequilibrium,guo2014probing,chase2022colored}. Our proposed choice of $\mathcal{H}(t)$ is advantageous for several reasons. Formally, it results in bounded and smooth variations of the random Hurst exponent. Furthermore, in the stationary state the PDF is given by the beta distribution (see~\cite{FITZ} and Section II in Supplemental Material~\cite{supp}),
\begin{align}
\label{eq-beta-dist}
    p(h)=\frac{(h-H_1)^{\lambda_{12}\tau-1}(H_2-h)^{\lambda_{21}\tau-1}}{(H_2-H_1)^{2\lambda\tau-1}B(\lambda_{21}\tau, \lambda_{12}\tau)},
\end{align}
where $H_1\leq h \leq H_2$, $\lambda=(\lambda_{12}+\lambda_{21})/2$ and $B(x, y)$ is the beta function. As can be seen from Eq.~(\ref{eq-beta-dist}), this distribution has four typical shapes (see Fig.~1 in the Supplemental Material~\cite{supp}), which ensures sufficient flexibility for different applications. Remarkably, the PDFs of the Hurst exponent extracted from  soft matter~\cite{wagner2017} and biological~\cite{sabri2020elucidating} experimental data were previously fitted with the bell-shaped unimodal beta distribution with $\lambda_{12}\tau,\lambda_{21}\tau >1$ \cite{balcerek2022fractional}. In what follows we basically restrict ourselves to such a shape, however for the sake of comparison we also study the bimodal case corresponding to $\lambda_{12}\tau,\lambda_{21}\tau <1$. Notably, a bimodal distribution of the Hurst exponents were reported in biological SPT experiments~\cite{han2020deciphering}.
The mean of $\mathcal{H}(t)$ is given by 
\begin{eqnarray}
\label{eq-mean-H}
   \langle \mathcal{H}(t) \rangle=\frac{H_1\lambda_{21}+H_2\lambda_{12}}{2\lambda}, 
\end{eqnarray}
while the ACVF is a combination of exponentials
(see, e.g.,~\cite{FITZ} and Section III in Supplemental Material~\cite{supp}) 
\begin{eqnarray}\label{gamma}
 &&\big\langle \big(\mathcal{H}(t)-\langle \mathcal{H}(t)\rangle\big) \big(\mathcal{H}(s)-\langle \mathcal{H}(s)\rangle\big)\big\rangle =  \nonumber \\
 &&\frac{\lambda_{12}\lambda_{21}(H_2-H_1)^2}{4\lambda^2(4\lambda^2\tau^2-1)}\left(2\lambda \tau e^{-|t-s|/\tau}-e^{-2\lambda |t-s|}\right).
\end{eqnarray}
We note that other choices of $\mathcal{H}(t)$, for instance the Ornstein-Uhlenbeck process or the squared Ornstein-Uhlenbeck process, also result in an exponential-like decay of the ACVF (see Section IX in Supplemental Material~\cite{supp}). However, unlike the smoothed telegraph process, for those choices ad-hoc boundary conditions need to be specified so that $\mathcal{H}$ remains bounded. Moreover, our choice is physically motivated due to the resultant stationary beta distribution of $\mathcal{H}$ and its flexibility to account for both uni- and bimodal distributions. 

In Fig.~\ref{fig:1}(a)  we show exemplary trajectories of $\mathcal{H}(t)$ while in Fig.~\ref{fig:1}(b)  we demonstrate the corresponding TeMBM trajectories. In addition, we present sample trajectories of FBM with three different Hurst exponents.  The simulation algorithms to generate the trajectories of  $\mathcal{H}(t)$ and $B_{\mathcal{H}}(t)$ are presented in Section IV in  Supplemental Material~\cite{supp}. The intermittent behaviour of the trajectory $B_{\mathcal{H}}(t)$  in the bimodal case is contrasted with that of the unimodal case. Indeed, it might be difficult to visually distinguish TeMBM in the unimodal case from FBM  (see, e.g., the blue curve vs. the pink curve in Fig.~\ref{fig:1}(b)).

FBM with random Hurst exponent (FBMRE) is a special case of MBM such that $\mathcal{H}(t)$ is constant for each trajectory but changes randomly from trajectory to trajectory. Such an approach is in the spirit of superstatistics~\cite{beck2001dynamical,beck2003superstatistics}. The properties of FBMRE with a beta distribution of the Hurst exponent were investigated in~\cite{balcerek2022fractional}. Apparently, the PDF and MSD of such FBMRE and TeMBM are the same for stationary $\mathcal{H}(t)$.  

While analysing stochastic time series, how can one distinguish TeMBM from the hierarchically lower level processes, namely, FBM and FBMRE, all of which exhibit power-law correlations?    Before addressing this issue we suggest a method that allows the estimation of the random Hurst exponent from the time averaged mean squared displacement (TAMSD).   We recall, for a time series $X=\{X_1,X_2,\dots, X_N\}$, where $X_i=X(t_i)$ are the observations recorded at time $t_i$, the TAMSD is defined as
\begin{eqnarray}
    \overline{\delta^2(\Delta)}=\frac{1}{N-\Delta}\sum_{j=1}^{N-\Delta}\left(X_{j+\Delta}-X_j\right)^2,
\end{eqnarray}
where $\Delta$ is the lag time. This widely-used observable is routinely measured, e.g., in SPT experiments \cite{metzler2014}.
In the case of FBM and FBMRE the TAMSD behaves as $\overline{\delta^2(\Delta)} \propto \Delta^{2H}$~\cite{deng2009ergodic,hubert2024riemann} where $H$ is the Hurst exponent of the trajectory for which the TAMSD is computed. Therefore, $H$ can be estimated as the slope of  the log-log plot of TAMSD vs. lag time.  By segmenting a trajectory into multiple segments with overlapping length, we extend this procedure also to obtain estimates of the Hurst exponent that changes along the trajectory (see Section V in Supplemental Material~\cite{supp}). Fig.~\ref{fig:2} validates this approach. More precisely, it shows that for both the unimodal (Fig.~\ref{fig:2}(a)) and bimodal (Fig.~\ref{fig:2}(b)) cases the distribution of estimated Hurst exponents from simulated TeMBM trajectories (``TeMBM'') agrees well with the  distribution (``Simulated'') of the ground truth values of Hurst exponents which generated the simulated trajectories, and also agrees with the analytical stationary distribution (``Analytical'').  

\begin{figure*}[t] 
\begin{subfigure}{.45\textwidth}
  \centering
  \includegraphics[width=.9\linewidth]{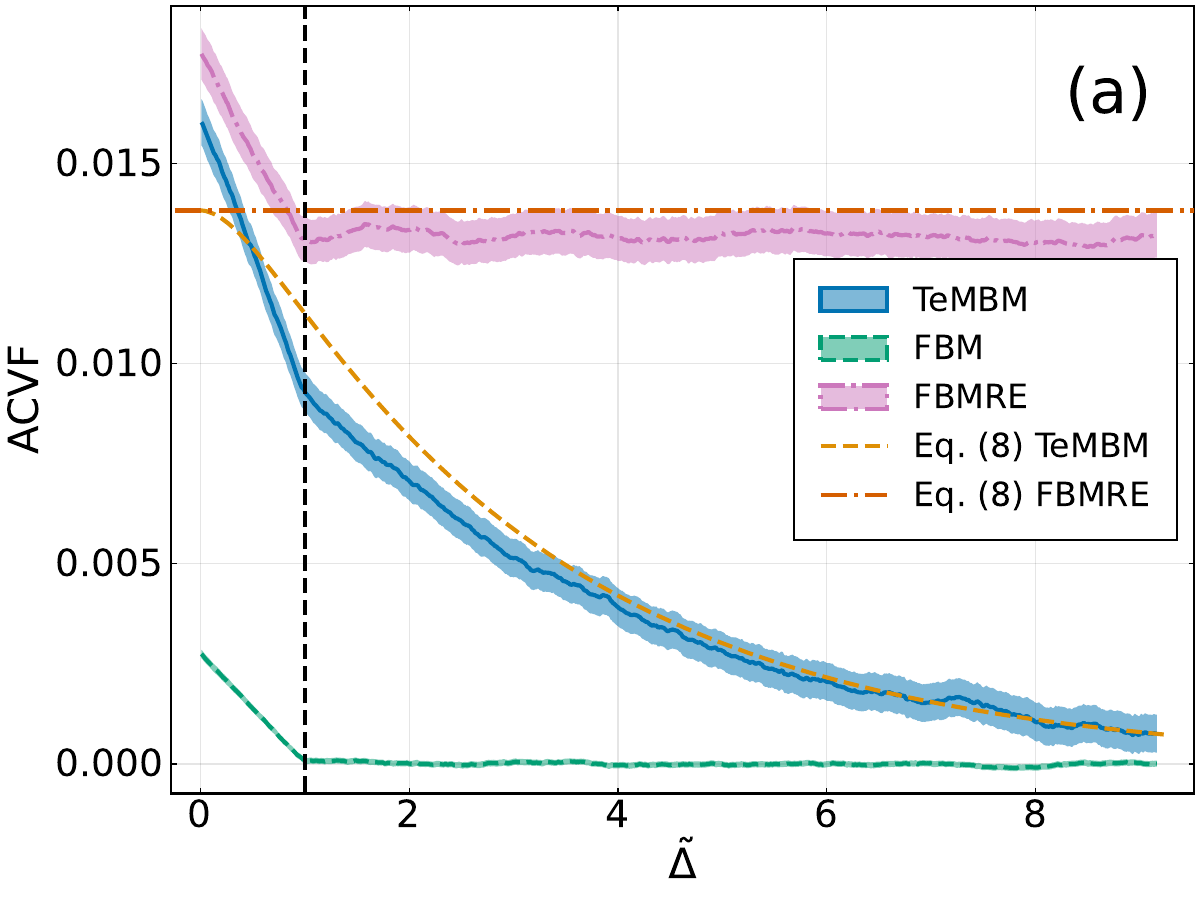}  
  \vskip -.2cm
\end{subfigure}
\begin{subfigure}{.45\textwidth}
  \centering
  \includegraphics[width=.9\linewidth]{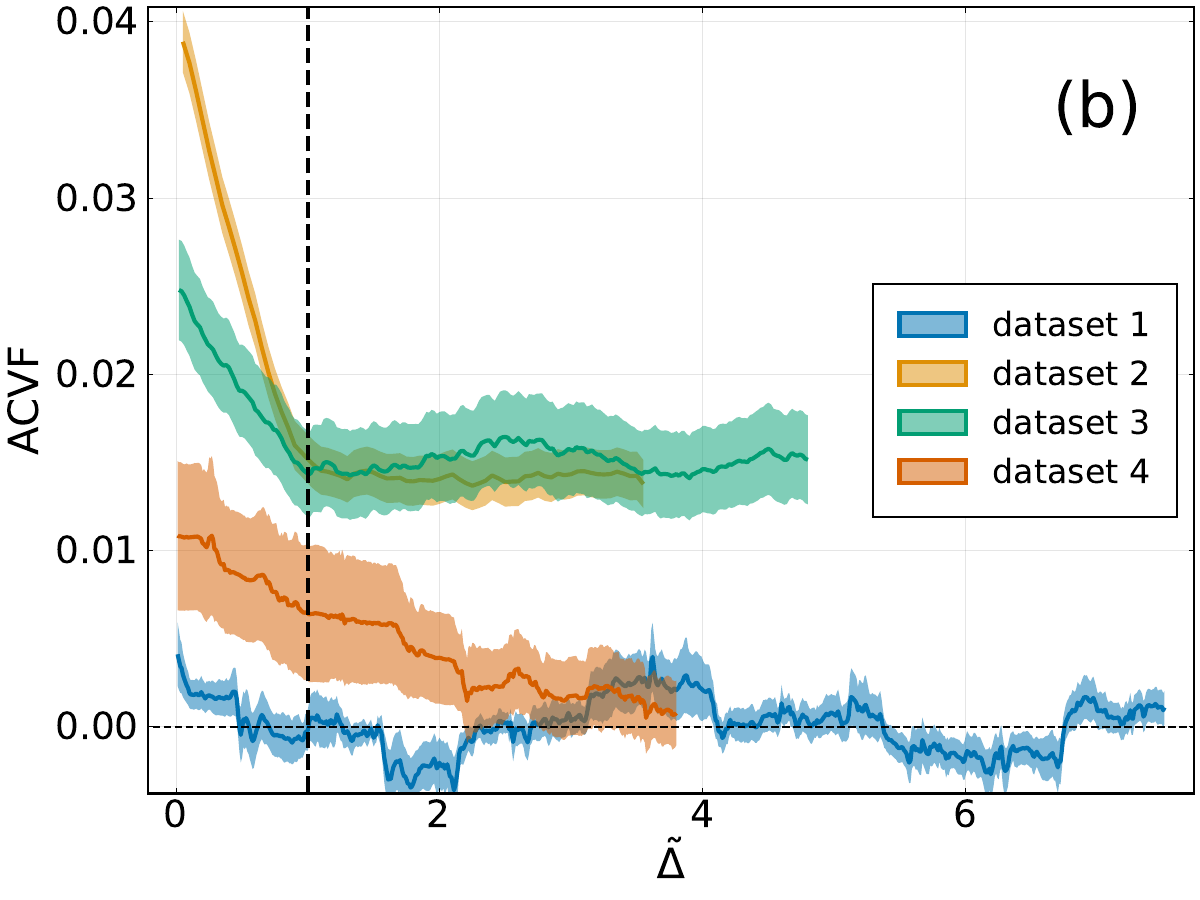}
  \vskip -.2cm
\end{subfigure}
\caption{(a) Sample ACVFs as function of the rescaled lag time $\tilde{\Delta}$ for the estimated Hurst exponent from $5,000$ trajectories of TeMBM (unimodal case), FBMRE (unimodal case) and FBM with $H=0.1$. The parameters of the processes are the same as used in Fig. \ref{fig:1} (see the figure caption for more details). Note that the analytical expression of ACVF for TeMBM is given by Eq.~(\ref{gamma}) which, upon setting $t=s$, also gives the analytical expression of ACVF for FBMRE with beta-distributed $H$. (b) Results of the process-distinguishing procedure for real data sets. The shaded regions in both (a) and (b) correspond to $95\%$ confidence intervals. See text for a description of the datasets.}
\label{fig:3}
\end{figure*}

To distinguish the processes we analyze the sample ACVF (calculated for sample trajectories) of the estimated Hurst exponents (see Section VI in Supplemental Material~\cite{supp}). For FBM,  the Hurst exponent is a constant value for each time point, resulting in a zero ACVF of the Hurst exponent. In the case of  FBMRE, for each trajectory the Hurst exponent takes on a random value thereby resulting in a constant, non-zero ACVF equal to the variance of $\mathcal{H}$.   
Finally, for TeMBM, the Hurst exponent is the smoothed telegraph process and the corresponding ACVF is given by a decaying function  (see Eq. (\ref{gamma})). Importantly, this classification procedure is applicable for discerning any  MBM process from FBMRE and FBM. 

Fig.~\ref{fig:3}(a) demonstrates the utility of this scheme for simulated TeMBM (unimodal case),  FBMRE (unimodal case) and FBM trajectories. The parameters are the same as in Fig.~\ref{fig:1}. The dashed black vertical line at $\tilde{\Delta}=1$ depicts the value of the lag time $\Delta$ normalized by the segment length used in the algorithm (see Section VI in Supplemental Material~\cite{supp}). We note that although segmentation is not required to estimate the Hurst exponent from FBM and FBMRE trajectories, in our analysis we follow the same procedure for all trajectories without \textit{a priori} assuming the underlying model. Indeed, as expected, the ACVF of the estimated Hurst exponents saturates at the zero-level for FBM, saturates at a constant, non-zero value for FBMRE, while it decays exponentially in the case of TeMBM for the chosen values of the parameters. Although the TAMSD-based method slightly underestimates the ACVF of $H$ for FBMRE and TeMBM, Fig.~\ref{fig:3}(a) shows that the simulation results are very close to the analytical expressions.

In Fig.~\ref{fig:3}(b) we present the results of the classification procedure applied to different experimental datasets. Dataset~1 consists of the time series of the cumulative sum of temperature anomalies obtained from mean daily temperature data, in the period from 1955-01-01 to 2020-12-31, collected at 10 different  meteorological stations in Germany~\cite{klein2002,web_temp}. Temperature anomalies are the deviations of the daily temperature at a given calendar day of the year from the average daily temperature at that particular calendar day, where the average is over all years considered, i.e., from 1955-2020~\cite{wallace2006atmospheric}. Dataset~2 are trajectories of quantum dots tracked in the cytoplasm of mammalian cells~\cite{sabri2020elucidating} and dataset~3 consists of trajectories of micron-sized beads tracked in mucin hydrogels at acidic conditions ($pH=2$) and with zero salt concentration~\cite{wagner2017}. Dataset~4 corresponds to the day-ahead electricity price in the year 2022 from the bidding zone between Germany and Luxembourg~\cite{web_electricity}. ``Bidding zone'' here refers to the largest area or region within which electricity producers and consumers submit their bids and offers without any technical constraints. We refer to Section VII in Supplemental Material~\cite{supp} for details on these datasets, and particularly how ensembles of trajectories are created in case of dataset~1 and dataset~4. Fig.~\ref{fig:3}(b) shows that for dataset~1 the sample ACVF of the estimated Hurst exponents is around zero, which we also observe for FBM trajectories in Fig.~\ref{fig:3}(a). This FBM-like behavior is consistent with previous analyses of such daily temperature data~\cite{massah2016,metzler2021largedev}. For  dataset~2 and dataset~3 one can see that the sample ACVFs for the estimated Hurst exponents stabilize at some non-zero levels, which indicates the correspondence of these datasets  to the FBMRE case. This is consistent with the results presented in~\cite{balcerek2022fractional} where we identified  beta distributions for the estimated Hurst exponents. For the electricity prices in dataset~4, we clearly see the decay of the sample ACVF of the estimated Hurst exponents, which indicates the multifractional case. In Fig.~\ref{fig:3}(b) we present the sample ACVF with the first observation on Wednesday. In Fig. 2 in Supplemental Material~\cite{supp} we show similar results for other starting points which highlights the robustness of the results with regard to the choice of the starting point.

One may ask why it is necessary to distinguish between the three classes of processes. 
With regards to SPT experiments, FBM with fixed Hurst exponent describes the dynamics of identical particles in homogeneous, viscoelastic media. FBMRE corresponds either to the dynamics of a heterogeneous ensemble of non-identical particles~\cite{cherstvy2018non,vacuoles2019} or the dynamics in a heterogeneous medium when each particle moves in a domain characterized by its own microstructure properties~\cite{weitz,Stiehl_2016,yael2021actin}. However, the most general cases, which can be mimicked by TeMBM, correspond to media or particle properties that vary in time, or to the situation when a tracked particle explores a large  spatial domain characterized by different Hurst exponents. TeMBM is sufficiently generic to describe the case of stochastically varying Hurst exponents while resulting in the observed beta distribution.   For experiments with identical particles, distinguishing between FBMRE and TeMBM allows us to separate fluctuations of the environment in time and in space.  

As for financial applications, estimates of the Hurst exponent often strongly suggest that a single parameter representing the long-term dependence is insufficient to capture the intricacies of the price evolution. The considerable variation in estimates can be succinctly explained by assuming that the degree of correlations undergoes fluctuations over time~\cite{bianchi2008multifractional, bianchi2014multifractional}. This contradicts the \textit{Efficient Market Hypothesis} (EMH), which within the paradigm of an ordinary Brownian setting, dictates that market prices incorporate all available information instantaneously~\cite{fama1970efficient,malkiel2003efficient}.  Indeed, as a consequence, this has led to the development of qualitative models such as \textit{Behavioural finance} ~\cite{camerer2003regulation,muradoglu2012behavioural}, which is based on the study of psychological influence on the behaviour of market practitioners, or \textit{Adaptive Market Hypothesis} ~\cite{lo2004adaptive} which relies on the concepts of evolutionary biology. Our proposed model could serve as a quantitative tool complementary to the qualitative models. It could allow the analytical assessment of how much the market prices deviate from EMH at any given time.  Fluctuations of $\mathcal{H}(t)$ at the same time could  describe the different market consequences and the investors' beliefs~\cite{bianchi2014multifractional}. Indeed $\mathcal{H}(t)$ can be understood as the weight assigned at a given time $t$ by an investor to past prices: $\mathcal{H}(t)=1/2$ indicates an efficient market, $\mathcal{H}(t)>1/2$ is indicative of a market whose future prices strongly depend on past prices and reacts slowly to new information (\textit{underreaction}) while $\mathcal{H}(t)<1/2$ denotes the belief that future prices will contradict the current prices and the market reacts strongly to new information (\textit{overreaction})~\cite{bianchi2014multifractional}.

Going back to physical systems, we note that TeMBM cannot account for all possible physical mechanisms and manifestations of heterogeneity. Our assumption that $\mathcal{H}(t)$ is a stationary process independent of $B(t)$ ensures that the resultant TeMBM is self-similar (see Theorem 4.1 in~\cite{ayachetaqqu2005}), but restricts its applicability to model ageing dynamics~\cite{weigel2011ergodic,tabei2013intracellular}. It is nevertheless an important step in the direction of research focused on the  comprehension and implications of heterogeneity in SPT experiments which started with the seminal articles on Brownian yet non-Gaussian diffusion~\cite{granick2009bng,granick2012bng} and the subsequent development of models with diffusing diffusivity~\cite{chubynsky2014dd,jain2016diffusion,aleks,tyagi2017non,lanoiselee2018diffusion,sposini2018random,wang2020fractional,wang2020unexpected,barkai2020packets,pastore2021rapid,sandalo2021length,pastore2022fickian,grossmann2024}. More advanced models should combine stochastic diffusivity  with stochastic Hurst exponent. Moreover, while in this Letter we consider a class of self-similar, power-law correlated processes, in light of the apparent heterogeneity in a wide variety of systems exhibiting anomalous diffusion, there is a need to generalize relevant anomalous diffusion models to include stochastic parameters~\cite{shlomi2024doubly,chen2024levy}. When establishing such generalizations, some care needs to be taken how the time dependence of the Hurst exponent and the diffusivity are incorporated, as shown for FBM-type processes with deterministic protocols~\cite{sabri2020elucidating,wang2023memory,slkezak2023minimal,balcerek2023modelling}. Concurrently, advanced data analysis methods such as those based on Bayesian statistics~\cite{krog2018,thapa2018,thapa2022} or machine learning~\cite{manzo2021,seckler2022bayesian,seckler2023machine} need to be developed to identify the best model given some empirical data.  

To summarize, we propose a generic, relatively simple analytical model of the multifractional process, namely telegraphic multifractional Brownian motion, that describes the temporal fluctuations of the system during its evolution. The Hurst exponent of this motion undergoes the smoothed telegraph process whose stationary PDF is given by the beta distribution. Such a choice is in agreement with Hurst exponent PDFs obtained in bio- and soft-matter experiments. We provide a methodology to distinguish between
three classes of power-law correlated random processes, namely FBM (with fixed Hurst exponent), FBM with Hurst exponent varying between different realizations, and telegraphic multifractional Brownian motion. The examples of the processes taken from biology, climate and finance illustrate the effectiveness of our approach. 

\section{Acknowledgments}
We thank Diego Krapf and Matthias Weiss for the dataset of quantum dots tracked in the cytoplasm of mammalian cells (dataset 2), and Caroline E. Wagner for the dataset of beads tracked in mucin hydrogels (dataset 3). The work of A.W. was supported by the National Science Centre, Poland, project 2020/37/B/HS4/00120. R.M. acknowledges funding from the German Ministry for Education and Research
(NSF-BMBF project STAXS). A.C. acknowledges funding from BMBF  project 01DK24006 PLASMA-SPIN-ENERGY.
\bibliography{bibliography}
\end{document}



\title{Supplemental Material for \\ Multifractional Brownian motion with telegraphic, stochastically varying exponent}
\author{Michał Balcerek}
\affiliation{Faculty of Pure and Applied Mathematics, Hugo Steinhaus Center, Wrocław University of Science and Technology, Wybrzeże Wyspiańskiego 27, 50-370 Wrocław, Poland}

\author{Samudrajit Thapa}
\email{thapa@pks.mpg.de }
 \affiliation{Max Planck Institute for the Physics of Complex Systems, Noethnitzer Straße 38, 01187 Dresden, Germany }
 \affiliation{Department of Physics, Indian Institute of Technology Guwahati, Guwahati 781039, Assam, India}
 
 \author{Krzysztof Burnecki}
 \affiliation{Faculty of Pure and Applied Mathematics, Hugo Steinhaus Center, Wrocław University of Science and Technology, Wybrzeże Wyspiańskiego 27, 50-370 Wrocław, Poland}
 
\author{Holger Kantz}
\affiliation{Max Planck Institute for the Physics of Complex Systems, Noethnitzer Straße 38, 01187 Dresden, Germany }
  
\author{Ralf Metzler}%
  \affiliation{Institute of Physics \& Astronomy, University of Potsdam,   14476 Potsdam, Germany}
  
\author{Agnieszka Wyłomańska}
\affiliation{Faculty of Pure and Applied Mathematics, Hugo Steinhaus Center, Wrocław University of Science and Technology, Wybrzeże Wyspiańskiego 27, 50-370 Wrocław, Poland}%

\author{Aleksei Chechkin}%
  \affiliation{Faculty of Pure and Applied Mathematics, Hugo Steinhaus Center, Wrocław University of Science and Technology, Wybrzeże Wyspiańskiego 27, 50-370 Wrocław, Poland}
  \affiliation{Institute of Physics \& Astronomy, University of Potsdam,   14476 Potsdam, Germany}
 \affiliation{German-Ukrainian Core of Excellence, Max Planck Institute of Microstructure Physics, \\ Weinberg 2,  06120 Halle, Germany}
 \affiliation{Akhiezer Institute for Theoretical Physics, National Science Center ‘Kharkiv Institute of Physics and Technology’, Akademichna
st.1, Kharkiv 61108, Ukraine}  

 \date{\today}  
\maketitle

\section{Derivation of the ACVF Eq. (2)}
Let us start with the spectral representation of FBM given by Eq.~(1) in the main text with $\mathcal{H}(t)=H=\mathrm{constant}$. The MSD is then given as
\begin{eqnarray}
   \langle B_H^2(t) \rangle &=&
    C^2(H) \int_{-\infty}^{\infty} \frac{\left(e^{\mathrm{i}\omega t}-1\right)\left(e^{-\mathrm{i}\omega t}-1\right)}{|\omega|^{2H+1}} d\omega =4C^2(H) \int_{0}^{\infty} \frac{1-\cos{\omega t}}{\omega^{2H+1}}d\omega \nonumber \\
    &=& \frac{\pi}{H\Gamma(2H)\sin(\pi H)}C^2(H) t^{2H} = \frac{2\pi}{\Gamma(2H+1)\sin(\pi H)}C^2(H) t^{2H}.
\end{eqnarray}

Thus, imposing $\langle B_H^2(t) \rangle=t^{2H}$ gives the expression for $C(H)$ as 
\begin{align}
    C(H)=\sqrt{\frac{\Gamma(2H+1)\sin(\pi H)}{2\pi}}.
\label{eq-CH} 
\end{align}
The autocovariance function (ACVF) for FBM calculated from the spectral representation thus takes the form
\begin{align}
  \langle B_{H}(t_1)B_H(t_2) \rangle = C^2(H) \int_{-\infty}^{\infty} \frac{\left(e^{\mathrm{i}\omega t_1}-1\right)\left(e^{-\mathrm{i}\omega t_2}-1\right)}{|\omega|^{2H+1}} d\omega=\frac{1}{2}\left[t_1^{2H}+t_2^{2H}-|t_2-t_1|^{2H} \right].
  \label{eq-fbm-autocorr1}
\end{align}

Let us consider Eq.~(1) of the main text and calculate the ACVF conditional on $\mathcal{H}(t)$: 
\begin{align}
    \langle B_{\mathcal{H}(t_1)}(t_1)B_{\mathcal{H}(t_2)}(t_2) \rangle =
    C(\mathcal{H}(t_1))C(\mathcal{H}(t_2)) \int_{-\infty}^{\infty} \frac{\left(e^{\mathrm{i}\omega t_1}-1\right)\left(e^{-\mathrm{i}\omega t_2}-1\right)}{|\omega|^{\mathcal{H}(t_1)+\mathcal{H}(t_2)+1}} d\omega.
    \label{eq-maska-corr1}
\end{align}
Now let's fix $t_1$ and $t_2$, and introduce $H=(\mathcal{H}(t_1)+\mathcal{H}(t_2))/2$. Then we can write (see Eq.~(\ref{eq-fbm-autocorr1}))
\begin{align}
\int_{-\infty}^{\infty} \frac{\left(e^{\mathrm{i}\omega t_1}-1\right)\left(e^{-\mathrm{i}\omega t_2}-1\right)}{|\omega|^{\mathcal{H}(t_1)+\mathcal{H}(t_2)+1}} d\omega &=\frac{1}{C^2(H)} \langle B_{H}(t_1)B_H(t_2) \rangle \nonumber \\  &=\frac{1}{2C^2(H)}\left[t_1^{2H}+t_2^{2H}-|t_2-t_1|^{2H} \right], 
\label{eq-integrand-maska-corr}
\end{align}
where $C(H)$ is given by Eq.~(\ref{eq-CH}). After plugging Eq.~(\ref{eq-integrand-maska-corr}) into 
Eq.~(\ref{eq-maska-corr1}) we get
\begin{eqnarray}
    \langle B_{\mathcal{H}(t_1)}(t_1)B_{\mathcal{H}(t_2)}(t_2) \rangle &=&\frac{C(\mathcal{H}(t_1))C(\mathcal{H}(t_2))}{2C^2(H)} \nonumber \\ &&\times \left[t_1^{\mathcal{H}(t_1)+\mathcal{H}(t_2)} +t_2^{\mathcal{H}(t_1)+\mathcal{H}(t_2)}-|t_2-t_1|^{\mathcal{H}(t_1)+\mathcal{H}(t_2)} \right].
\end{eqnarray}
Thus we arrive at Eq.~(2) in the main text.

\section{Beta distribution as the stationary PDF for smoothed telegraph process}
Let us obtain the stationary PDF of the process governed by the Langevin equation
\begin{align}
    \frac{dx}{dt}=f(x)+\xi(t),
   \label{eq-langevin1} 
\end{align}
where $f$ is a deterministic function and $\xi(t)$ is a telegraph process, i.e. a stationary dichotomic Markov process that jumps between two values $c_1$ and $c_2$, $c_1<c_2$,  with mean rates $\lambda(c_1 \rightarrow c_2)=\lambda_{12}$ and $\lambda(c_2 \rightarrow c_1)=\lambda_{21}$.  The mean and the ACVF of $\xi$ are given respectively by
\begin{align}
    \langle \xi \rangle=\frac{1}{2\lambda}\left(\lambda_{12}c_2+\lambda_{21}c_1\right)
\end{align}
and
\begin{align}
    \langle \xi(t)\xi(t') \rangle=\langle \xi \rangle^2+\frac{\lambda_{12}\lambda_{21}(c_2-c_1)^2}{4\lambda^2}\exp{\left(-2\lambda|t-t'|\right)},
\end{align}
where $\lambda=(\lambda_{12}+\lambda_{21})/2$.

Here we employ the approach and notations used in ~\cite{sancho1984}. Let's introduce the ``microscopic density'',
\begin{align}
    \rho(x,t)=\delta(x-x(t)),
\end{align}
which obeys the equation
\begin{align}
    \frac{\partial \rho}{\partial t}+\frac{\partial}{\partial x}\left[ (f(x)+\xi(t)) \rho \right]=0.
    \label{eq-rho}
\end{align}
Introducing $p(x,t)=\langle \delta(x-x(t)) \rangle$ and averaging Eq.~(\ref{eq-rho}) one gets
\begin{align}
    \frac{\partial p}{\partial t}+\frac{\partial}{\partial x}(f(x)p)+\frac{\partial p_1}{\partial x}=0,
    \label{eq-p}
\end{align}
where $p_1(x,t)= \langle \xi(t)\rho \rangle$. Following~\cite{sancho1984} and using the ``formula of differentiation''~\cite{shapiro1978}, we arrive at
\begin{align}
    \frac{\partial p_1}{\partial t}=-2\lambda p_1-\frac{\partial}{\partial x}\left[ f(x)p_1\right]-(c_1+c_2)\frac{\partial p_1}{\partial x}+c_1c_2\frac{\partial p}{\partial x}+(\lambda_{12}c_2+\lambda_{21}c_1)p.
    \label{eq-p1}
\end{align}

Now we consider $f(x)=-\gamma x$ where $\gamma$ is constant, and after some transformations of Eqs.~(\ref{eq-p}) and (\ref{eq-p1}) we arrive at the following closed equation for $p$ 
\begin{align}
    \frac{\partial^2 p}{\partial t^2}&+(2\lambda-3\gamma)\frac{\partial p}{\partial t} +(c_1+c_2-2\gamma x)\frac{\partial^2p}{\partial x \partial t}+\left[\gamma^2x^2+c_1c_2-\gamma x(c_1+c_2) \right]\frac{\partial^2 p}{\partial x^2}  \nonumber \\
    & + \left[4\gamma^2x-2\lambda\gamma x-2\gamma(c_1+c_2)+\lambda_{12}c_2+\lambda_{21}c_1\right]\frac{\partial p}{\partial x}+2\gamma(\gamma-\lambda)p=0.
    \label{eq-p-closed-full}
\end{align}
We are looking for the stationary solution of Eq.~(\ref{eq-p-closed-full}). Putting $\frac{\partial p}{\partial t}=0$, and then integrating with respect to $x$ we obtain the particular solution as
\begin{align}
    p(x)=\mathcal{N}\left(x-\frac{c_1}{\gamma}\right)^{\frac{\lambda_{12}}{\gamma}-1}\left(\frac{c_2}{\gamma}-x\right)^{\frac{\lambda_{21}}{\gamma}-1}, \quad \frac{c_1}{\gamma}<x<\frac{c_2}{\gamma},
\end{align}
where $\mathcal{N}$ is the normalization constant.
To go back to our notations in the main text, we change $c_{1,2}/\gamma \rightarrow H_{1,2}$, $\gamma \rightarrow \tau^{-1}$ and arrive at the beta distribution presented in Eq.~(6) of the main text.

\begin{figure}[h!]
\centering	
\includegraphics[scale=0.6]{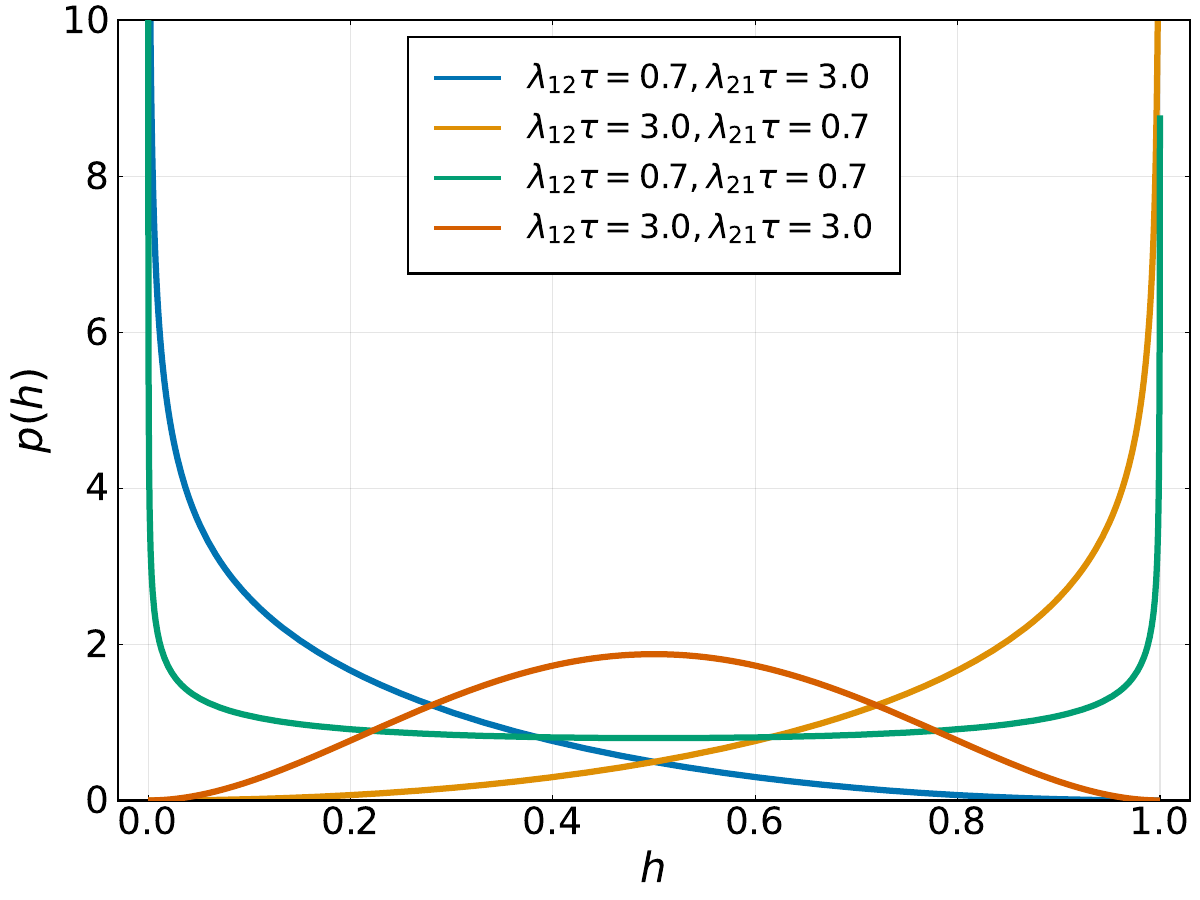}
\caption{Representative shapes of the beta distribution (Eq.~(6) in the main text) highlights its flexibility via specific choice of parameters to potentially describe  various experimental observations. }
\label{fig-betadist_shapes}
\end{figure}

Fig. \ref{fig-betadist_shapes} shows how the flexibility of beta distribution can be leveraged to realize different shapes depending on the choice of parameters.  

\section{Derivation of the ACVF  of smoothed telegraph process}
A simple derivation of Eqs.~(7) and (8) in the main text is specified in~\cite{FITZ}. Because of stationarity, $\langle \mathcal{H}(t) \rangle= \langle \mathcal{H}_{TP}(t) \rangle$, and in notations of~\cite{FITZ} it is given by
\begin{align}
\mu_{x1}=q_0a_0 + q_1a_1 ,
\end{align}
while ACVF is given by Eq.(12) in~\cite{FITZ}. To establish the correspondence with Eq.~(8) in the main text, one changes
\begin{align}
q_0 = \nu_1/\nu \rightarrow \lambda_{21}/(2\lambda),
\; q_1 = \nu_0/\nu \rightarrow \lambda_{12}/(2\lambda) ,
\; a_0 \rightarrow H_1 , \; a_1 \rightarrow H_2 .
\end{align}

\section{Alternative models of $\mathcal{H}(t)$}
The Ornstein-Uhlenbeck process (OUP) given as a stationary solution of the stochastic differential equation
\begin{align}
    dX_t = \theta (\mu - X_t)dt + \sigma dB_t
\end{align}
and the square of the OUP are good alternatives to model $\mathcal{H}(t)$. Similarly to the smoothed telegraph process, they can model the time-varying nature of $\mathcal{H}(t)$, with the advantage of being mean-reverting and having well-understood stochastic properties. On top of that, we know that the OUP is a continuous-time Gaussian process with an exponentially decaying autocorrelation function. When squared, it provides non-negative values important for considering the range of $\mathcal{H}(t)$. Naturally, such models have to be treated with greater care than the smoothed telegraph process, since their possible values span all reals (for OUP) and non-negative reals (for its square). One alternative is to utilize the OUP reflected at the origin~\cite{skorokhod1961stochastic, giorno1986some, ward2003diffusion}.

\begin{figure*}[h!] 
  \centering
  \includegraphics[width=.6\linewidth]{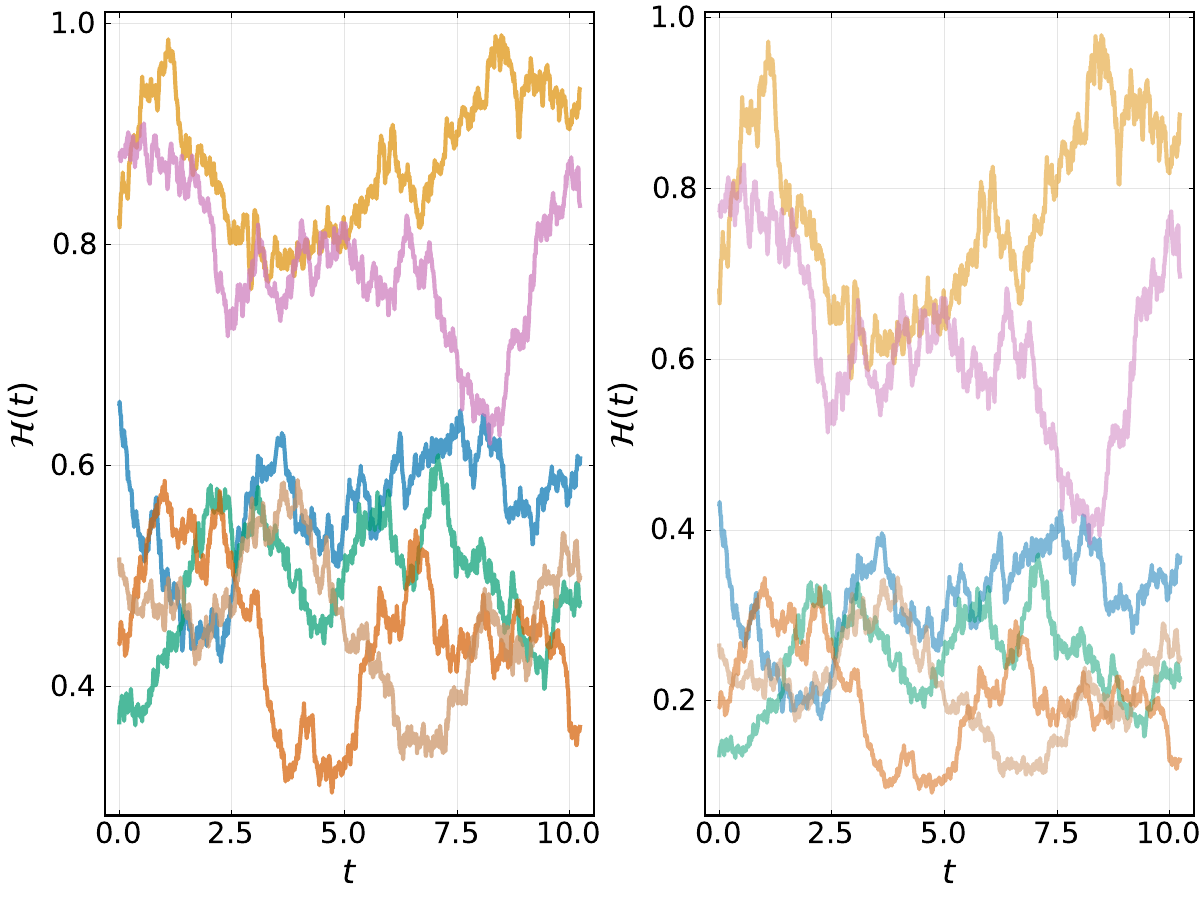}
  \vskip -.2cm
  \caption{Sample paths of the OUP with parameters $\theta = 0.08, \mu = 0.5, \sigma = 0.06$ (left panel) and its square (right panel).}
  \label{fig:sm_oucir_paths}
\end{figure*}

\begin{figure*}[h!] 
  \centering
  \includegraphics[width=.6\linewidth]{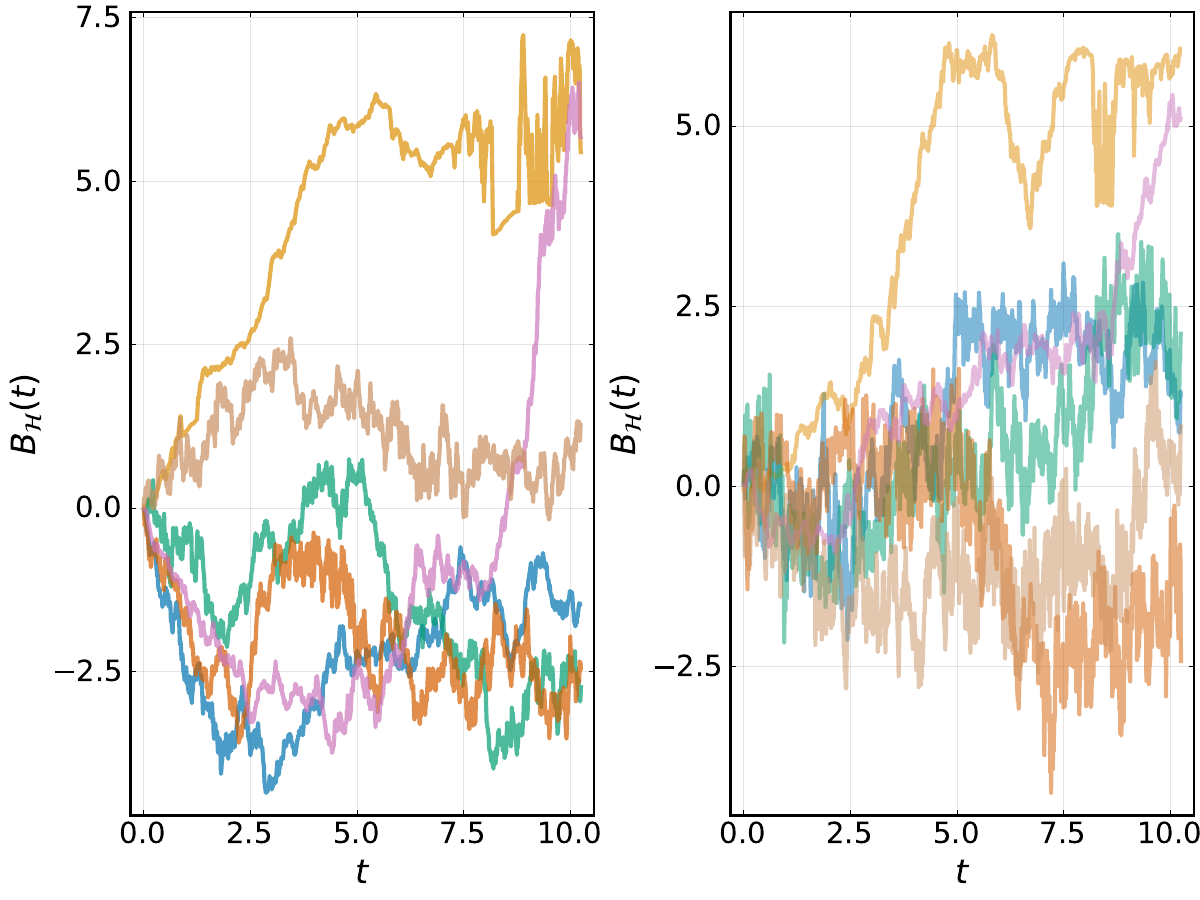}
  \vskip -.2cm
  \caption{Sample paths of MBM with $\mathcal{H}(t)$ modelled by the OUP (left panel) and by its square (right panel). The parameters of the OUP are the same as before, i.e., $\theta = 0.08, \mu = 0.5, \sigma = 0.06$.}
  \label{fig:sm_mbm_oucir}
\end{figure*}

\begin{figure*}[h!] 
  \centering
  \includegraphics[width=.5\linewidth]{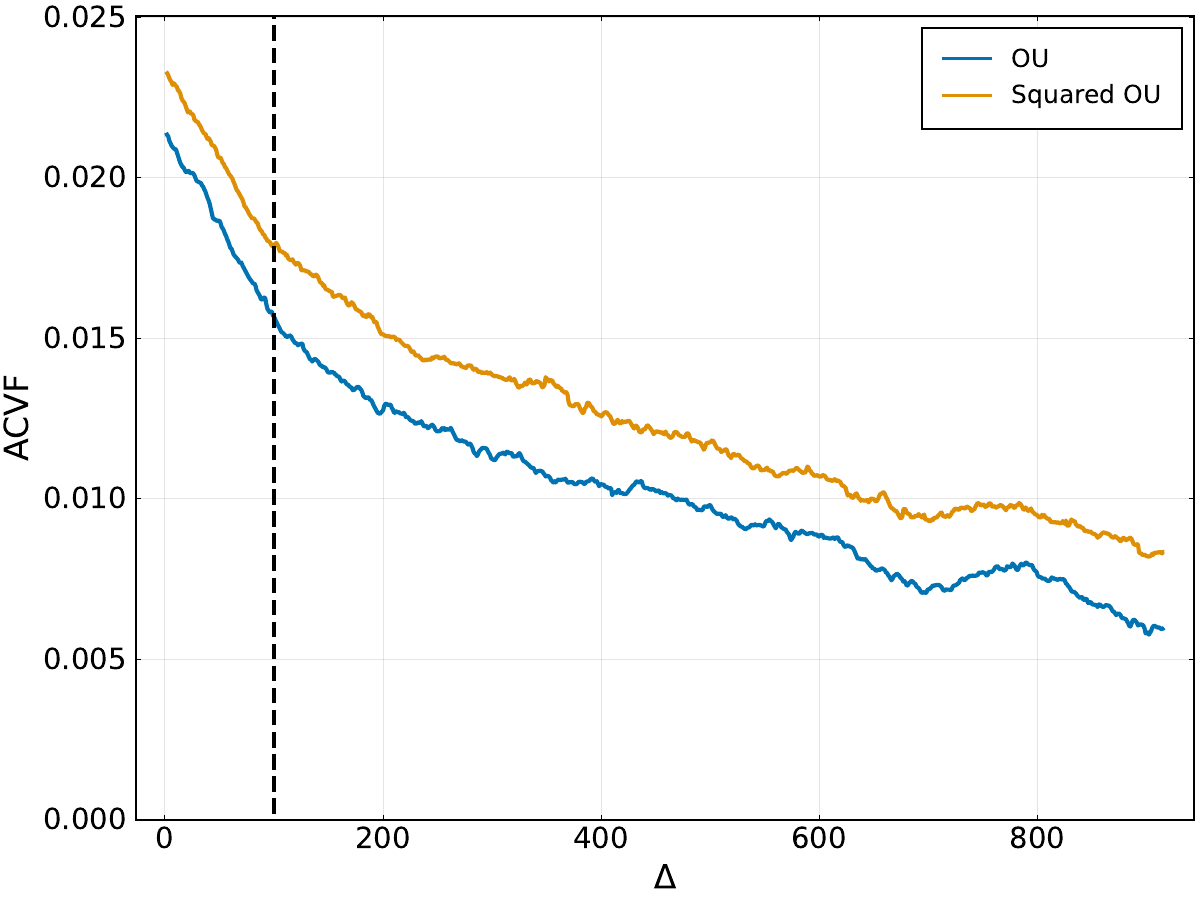}
  \vskip -.2cm
  \caption{Sample ACVFs as function of the lag time ${\Delta}$ for the estimated Hurst exponents from $5{,}000$ trajectories of MBM with $\mathcal{H}(t)$ modelled by the OUP (``OU'') and by its square (``Squared OU''). The parameters of the OUP are the same as before, i.e., $\theta = 0.08, \mu = 0.5, \sigma = 0.06$.}
  \label{fig:sm_oucir}
\end{figure*}

In Fig.~\ref{fig:sm_oucir_paths} we present the behavior of sample realizations of $\mathcal{H}(t)$, where the left panel corresponds to the OUP with parameters $\theta = 0.08, \mu = 0.5, \sigma = 0.06$ and time-step 0.01, and the right panel corresponds to its square.
In Fig.~\ref{fig:sm_mbm_oucir} we present trajectories of MBM that correspond to the chosen model of $\mathcal{H}(t)$ -- the left panel utilizes the OUP model for $\mathcal{H}(t)$, the right panel corresponds to its square. In Fig.~\ref{fig:sm_oucir} we show sample ACVFs as functions of lag time $\Delta$ for the estimated Hurst exponents of the two processes considered here. They both exhibit an exponential-like decay similarly to the ACVF of the smoothed telegraph process.

\section{Simulation algorithm}
In this part we describe in details the simulation algorithm for \textbf{telegraphic multifractional Brownian motion}  $B_{\mathcal{H}}(t)$ in times $t_1 < t_2 < \ldots < t_n$. For the sake of simplicity let's consider equally spaced $t_k$'s, i.e. $t_k = k\cdot \Delta, k = 1, 2, \ldots, n$. Then the algorithm to simulate TeMBM is as follows:
\begin{enumerate}
    \item Simulate the trajectory of the process $\mathcal{H}(t)$  in times $t_1, t_2, \ldots, t_n$. The algorithm depends on the type of the process. Below we present how to simulate the sample trajectory of the smoothed telegraph process. 
    \item Given $\mathcal{H}(t_k)$ for $k=1, 2, \ldots, n$ construct the autocovariance matrix of vector \\
    $[B_{\mathcal{H}}(t_1), B_{\mathcal{H}}(t_2), \ldots, B_{\mathcal{H}}(t_n)]'$, that is
    \begin{align*}
        \Sigma = [\langle B_{\mathcal{H}}(t_i) B_{\mathcal{H}}(t_j) \rangle]_{1\leq i, j \leq n},
    \end{align*}
    where $\langle B_{\mathcal{H}}(t_i) B_{\mathcal{H}}(t_j) \rangle$ 
    is the ACVF of the MBM given in Eq. (2) in the main text.
    \item Use Cholesky algorithm, i.e. decompose matrix $\Sigma$ to find the lower triangular matrix $L$, and then 
    \begin{align*}
        \begin{bmatrix}
            B_{\mathcal{H}}(t_1)\\
            B_{\mathcal{H}}(t_2)\\
            \vdots\\
            B_{\mathcal{H}}(t_n)
        \end{bmatrix} = 
        L \cdot \begin{bmatrix}
            Z_1\\
            Z_2\\
            \vdots\\
            Z_n
        \end{bmatrix}
    \end{align*}
    where $Z_1, Z_2, \ldots, Z_n$ are independent identically distributed  standard normal random variables. For example, it can be done using a built-in function \texttt{chol(}$\Sigma$\texttt{, 'lower')} in Matlab; \texttt{np.linalg.cholesky(}$\Sigma$\texttt{)} in Python using \texttt{numpy} library; or \texttt{cholesky(}$\Sigma$\texttt{).L} from \texttt{LinearAlgebra} library in Julia. 
        
\end{enumerate}
Simulation algorithm for \textbf{smoothed telegraph process} $\mathcal{H}(t)$ in times $t_1 < t_2 < \ldots < t_n$ is as follows:
\begin{enumerate}
    \item First, generate telegraph process $\mathcal{H}_{TP}(t)$ in the same times $t_1, t_2, \ldots, t_n$:
        \begin{enumerate}
            \item Set current state \texttt{state = rand(0,1)}, i.e. a random number 0 or 1. If a stationary version of the telegraph process is required, choose 0 with probability $\frac{\lambda_{12}}{\lambda_{12} + \lambda_{21}}$, and 1 with probability $\frac{\lambda_{21}}{\lambda_{12} + \lambda_{21}}$.
            \item Set \texttt{ind = 1}
            \item If \texttt{state == 0} then set \texttt{len =} $\lceil$\texttt{rand($\mathcal{E}_0$)}$\rceil$, a random number from exponential distribution with rate $\Delta\cdot \lambda_{12}$, otherwise \texttt{len =} $\lceil$\texttt{rand($\mathcal{E}_1$)}$\rceil$, a random number from exponential distribution with rate $\Delta\cdot \lambda_{21}$.
            \item Set $\mathcal{H}_{TP}(t_\texttt{ind}), \mathcal{H}_{TP}(t_{\texttt{ind}+1}), \ldots, \mathcal{H}_{TP}(t_\texttt{ind+len-1})$ to \texttt{state}.
            \item \texttt{state = 1 - state}
            \item \texttt{ind = ind + len}
            \item If \texttt{ind} $< n$ then go back to (c).
            \item Rescale the values of $\mathcal{H}_{TP}$ to $H_1$ and $H_2$ instead of $0$ and $1$:
            $\mathcal{H}_{TP}(t_k) = \mathcal{H}_{TP}(t_k)\cdot(H_2-H_1)+H_1$ for $k = 1, 2, \ldots, n$.
        \end{enumerate}
    \item To obtain smoothed telegraph process:
        \begin{enumerate}
            \item Set $\mathcal{H}(t_1) = \texttt{rand(}\mathcal{B}\texttt{)}$, a random number from stationary STP distribution given by beta distribution in Eq. (4) in the main text.
            \item For $k=2, 3, \ldots, n-1$ set $\mathcal{H}(t_{k+1}) = \mathcal{H}(t_{k}) + \Delta\cdot \frac{\mathcal{H}_{TP}(t_{k}) - \mathcal{H}(t_{k})}{\tau}.$
        \end{enumerate}
\end{enumerate}

\section{Estimation algorithm}
\label{sec:est-alg}
Let us consider $M$ random samples of length $N$ 
\begin{align}\label{X}\mathbb{X}_N^i=\{X^i(t_1),X^i(t_2),\ldots,X^i(t_N)\}, ~~~i=1,2,\ldots,M.\end{align} The estimation algorithm for the Hurst exponent is as follows:
\begin{enumerate}
    \item We select a segment length $w$ and an overlapping length $o$.
    \item Each sample trajectory $\mathbb{X}_N^i$, $i=1,2\ldots,M$ is divided into segments of length $w$ with the overlapping length of $o$ points. We denote them as $\mathbb{X}_N^{i,1}, \mathbb{X}_N^{i,2},\ldots,\mathbb{X}_N^{i,m}$, where $m$ is the total number of segments constructed from the $i$-th trajectory.
    \item For each segment $\mathbb{X}_N^{i,j}$ we use TAMSD-based approach to estimate the Hurst exponent. The estimated values are denoted as ${H}^{i,j}$, $i=1,2,\ldots,M$, $j=1,2,\ldots,m$.
   \end{enumerate}

\section{Distinguishing algorithm}
We consider $M$ random samples of length $N$, see Eq. (\ref{X}) for the notation.
To discern whether the sample trajectories correspond to  FBM, FBMRE or TeMBM  we proceed as follows:
\begin{enumerate}
       \item For each sample trajectory $\mathbb{X}_N^i$, $i=1,2\ldots,M$ we estimate the Hurst exponents ${H}^{i,j}$, $i=1,2,\ldots,M$, $j=1,2,\ldots,m$, according to the  estimation algorithm presented above. 
    \item For each $j,k=1,2,\ldots,m$ we estimate the sample ACVF  
    \begin{eqnarray}
        {\gamma}(j,k)={\sum_{i=1}^M\left({H}^{i,j}-\overline{{H}^{j}}\right)\left({H}^{i,k}-\overline{{H}^{k}}\right)},
    \end{eqnarray}
    where $\overline{{H}^{j}}=\frac{1}{M}\sum_{i=1}^M{H}^{i,j}$.
    \item For fixed $j$ we analyze the function  ${\gamma}(j,k)$. More precisely, if for large values of $k$  ${\gamma}(j,k)$ stabilizes at zero level, then  the sample trajectories correspond to FBM. On the other hand, if  ${\gamma}(j,k)$ stabilizes at some non-zero level, then the sample trajectories correspond to FBMRE. Finally, if we observe a decay of ${\gamma}(j,k)$, the sample trajectories can be attributed to TeMBM. 
\end{enumerate}
Let us note, that in the above-described distinguishing algorithm, we do not take into account the diffusion coefficient. This can be done only under certain conditions which we discuss here. The Hurst index ${H}$ is estimated from the slope of the log-log plot of  TAMSD vs. $\Delta$ after selecting a window size, $w$. The diffusion-coefficient appears only as the $y$-intercept in the log-log plot, and therefore should not significantly effect the estimation of ${H}$. Fig.~2 in the main text shows that our estimates of ${H}$ are reasonable and results in the same distribution as the original stationary distribution of $H$ that underlay the generation of TeMBM trajectories.     

The Python source code for the simulations and estimations is available on GitHub \cite{web:source}.

\section{Description of the real-world datasets}
\textbf{Dataset~1:} This dataset consists of the time series of mean daily temperature data, from 01.01.1955 to 31.12.2020, collected at 10 different  meteorological stations in Germany~\cite{klein2002,web_temp}. The stations were chosen such that there are no missing data points. Exact details of the stations are provided in the Table~\ref{tab:meteo}. We consider the time series of temperature anomalies, namely, the deviations of the daily temperature at a given calendar day of the year from the average daily temperature at that particular calendar day, where the average is over all the years considered. It was shown previously that the time series of such temperature anomalies could be modelled with fractional Gaussian noise but with additional short range correlations of 4-5 days~\cite{massah2016,thapa-largedev}. In order to remove the short range correlations we take a weekly average of the temperature anomalies, and then take a cumulative sum to  obtain FBM-like trajectories. Finally to create a larger ensemble of trajectories, we split each trajectory into 4 trajectories, so that our dataset consists of 40 trajectories, each with $N=860$ data points. 

\begin{table}[h]
    \centering
    \caption{Details of the meteorological stations from which the daily mean temperature data was taken to construct dataset~1.}

    \begin{tabular}{|c|c|c|c|}
        \hline
        Location & Latitude & Longitude & Height  \\
          &(degrees:minutes:seconds)  & (degrees:minutes:seconds)  & (meters)\\
        \hline
        Bamberg & +49:52:31 & +010:55:18 & 240\\
        \hline
        Berlin-Dahlem & +52:27:50 & +013:18:06  & 51\\
        \hline
        Bremen & +53:02:47 & +008:47:57 &    4 \\
        \hline
        Frankfurt & +50:02:47 & +008:35:54 &  112 \\
        \hline
        Hohenpeissenberg & +47:48:06 & +011:00:42 &  977 \\
        \hline
        Jena Sterwarte & +50:55:36 & +011:35:03 &  155 \\
        \hline
        Muenchen & +48:09:51 & +011:32:39 &  515 \\
        \hline
        Potsdam & +52:23:00 & +013:03:50 &   81 \\
        \hline
       Schwerin  & +53:38:39 & +011:23:18 &   59 \\
       \hline
       Zugspitze & +47:25:19 & +010:59:12 & 2964 \\
       \hline
    \end{tabular}
    \label{tab:meteo}
\end{table}

\textbf{Dataset~2:} This dataset consists of 3834 trajectories---each with $N=100$ data points measured with an experimental time resolution of $100$ ms---corresponding to quantum dots tracked in the cytoplasm of mammalian cells~\cite{sabri2020elucidating}. It was shown previously that this dataset corresponds to FBMRE with the Hurst exponent beta distributed~\cite{balcerek2022fractional}.

\textbf{Dataset~3:} This dataset consists of 532 trajectories---each with $N=300$ data points measured with an experimental time resolution of $33$ ms--- corresponding to micron-sized beads tracked in mucin hydrogels at acidic conditions ($pH=2$) and with zero salt concentration~\cite{wagner2017}. It was shown previously that this dataset too corresponds to FBMRE with the Hurst exponent beta distributed~\cite{cherstvy2019non,balcerek2022fractional}.

\textbf{Dataset~4:} This dataset describes day ahead electricity price in the year 2022 from the bidding zone between Germany and Luxembourg (BZN|DE-LU). The data are quoted every 15 minutes and are publicly available on the web page \cite{web_electricity}. In order to obtain an ensemble of trajectories we split the time series into 51 trajectories corresponding to week periods (Monday 00:00 -- Sunday 23:45). Consequently, each trajectory has 672 observations. 

\section{Additional results for electricity price data}
\begin{figure*}[h!] 
  \centering
  \includegraphics[width=.5\linewidth]{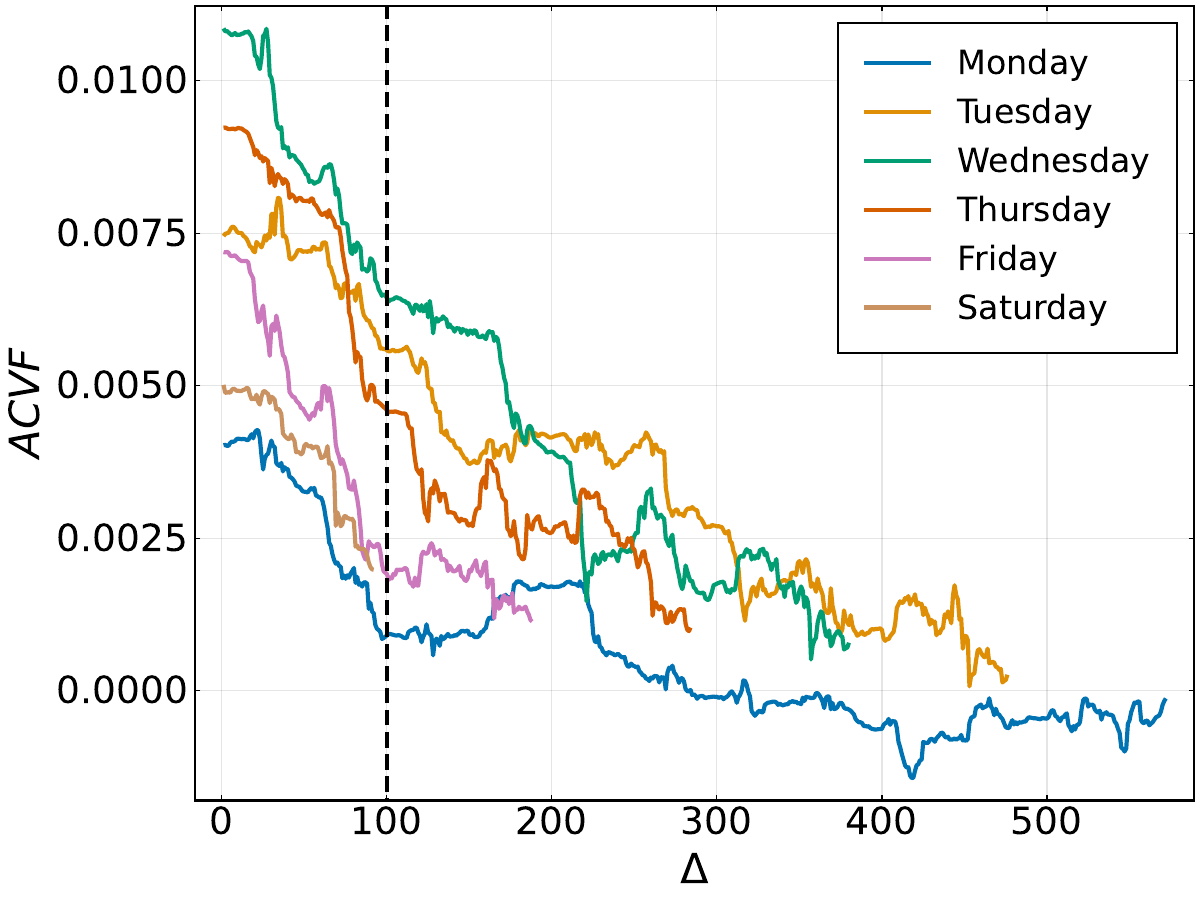}
 \vskip -.2cm
  \caption{Results of the process-distinguishing procedure for electricity price data for different starting points $t$.  The labels ``Monday'', ``Tuesday'', etc. mean that the starting point $t$ in sample ACVF ${\gamma}(t,t+\Delta)$ calculation is the first observation (hour) on Monday, Tuesday, etc., respectively.}
  \label{fig:5d}
\end{figure*}

In Fig.~\ref{fig:5d} we demonstrate the results for six different starting time points $t$ used in the distinguishing procedure. 
The labels ``Monday'', ``Tuesday'', etc. mean that the starting point $t$ in sample ACVF ${\gamma}(t,t+\Delta)$ calculation is the first observation (hour) on Monday, Tuesday, etc., respectively. All the cases are indicative of TeMBM. 

\clearpage

\bibliographystyle{apsrev4-1}
\bibliography{bibliography}